\documentclass[a4paper,11pt]{article}
\pdfoutput=1
\usepackage{jcappub}

\usepackage[normalem]{ulem}

\usepackage{amssymb} %maths
\usepackage{amsmath} %maths
\usepackage{bm} %maths
\usepackage{esint} %maths
\usepackage[utf8]{inputenc} %useful to type directly 
\usepackage{enumerate}
\usepackage{graphicx}
\usepackage{ifthen}
\usepackage{xfrac}
\usepackage{hyperref}

\newcommand{\be}{\begin{equation}}
\newcommand{\ee}{\end{equation}}
\newcommand{\nn}{\nonumber\\}
\newcommand{\alp}{\alpha'}
\newcommand{\llang}{\langle \! \langle}
\newcommand{\rrang}{\rangle \! \rangle}

\title{\boldmath Graviton propagation within the context of the D-material universe}

\vspace{1cm}
\author{Thomas Elghozi,}
\author{Nick E. Mavromatos,}
\author{Mairi Sakellariadou}

\affiliation{Theoretical Particle Physics and Cosmology Group, Department of Physics,\\ King's College London, University of London, Strand, London WC2R 2LS, United Kingdom}

\emailAdd{thomas.elghozi@kcl.ac.uk}
\emailAdd{nikolaos.mavromatos@kcl.ac.uk}
\emailAdd{mairi.sakellariadou@kcl.ac.uk}

\date{\today}

\abstract{Motivated by the recent breakthrough of the detection of Gravitational Waves (GW) from coalescent black holes by the aLIGO interferometers, we study the propagation of GW in the {\sl D-material universe}, which we have recently shown to be compatible with large-scale structure and inflationary phenomenology. The medium of D-particles induces an effective mass for the graviton, as a consequence of the formation of recoil-velocity field condensates due to the underlying Born-Infeld dynamics. There is a competing effect, due to a super-luminal refractive index, as a result of the gravitational energy of D-particles acting as a dark matter component, with which propagating gravitons interact. We examine conditions for the condensate under which the latter effect is sub-leading. We argue that if quantum fluctuations of the recoil velocity are relatively strong, which can happen in the current era of the universe, then the condensate, and hence the induced mass of the graviton, can be several orders of magnitude larger than the magnitude of the cosmological constant today.
Hence, we constrain the graviton mass using aLIGO and pulsar timing observations (which give the most stringent bounds at present). In such a sub-luminal graviton case, there is also a gravitational Cherenkov effect for ordinary high energy cosmic matter, which is further constrained by means of ultra-high-energy cosmic ray observations. Assuming cosmic rays of extragalactic  origin, 
the bounds on the quantum condensate strength, based on the gravitational Cherenkov effect, are of the same order as those from aLIGO measurements, in contrast to the case where a galactic origin of the cosmic rays is assumed, in which case the corresponding 
bounds are much weaker.}

\begin{document}
\begin{flushleft}
KCL-PH-TH/2016-24\\
LCTS/2016-16
\end{flushleft}

\maketitle
\flushbottom

%%%%%%%%%%%%%%%%%%%%%%%%%%%%%%%%%%%%%%%%%%%%%%%%%%%%%%%%%%
\section{Introduction}

The Gravitational Wave (GW) signal GW150914 detected by Advanced LIGO (aLIGO)~\cite{LIGO}, based on the effects of the arrival of GW on the arms of the pertinent interferometric devices due to the distortion of the neighbouring space-time, opened a new window on the fundamental laws governing our universe. The foreseen extended network of terrestrial interferometers combined with eLISA, the first GW observatory in space, may eventually detect even quantum aspects of gravity, or at least falsify quantum gravity models which entail Lorentz Invariance Violation (LIV) for which there are already stringent restrictions from various sources~\cite{LIV}.

A microscopic LIV model which evades such constraints is the D-material universe~\cite{dfoam,emsy}, a brane-world (viewed as our three-spatial dimensional universe) propagating in a higher-dimensional bulk populated by D-particle stringy defects. Depending on the type of string theory considered, these defects can be either point-like or compactified higher-dimensional 3-branes wrapped around three cycles, thus appearing from the point of view of an observer on the brane world as effectively ``point-like'' defects. The interaction, for instance of a photon with the population of such D-particles, crossing or being confined on our brane-world, leads to time delays proportional to the energy of the incident photon. This effectively yields a linear modification of the corresponding dispersion relation, suppressed though, not by the Planck scale but by an effective mass scale inversely proportional to the linear density $n^*(z)$ of the defects encountered in the path of the photon~\cite{delays}:
\begin{equation} \label{stringdr}
	E = p \left( 1 - \frac{p}{M_{\rm QG}} \right) ~\mbox{where} \quad M_{\rm QG} = \frac{M_{\rm Pl}}{n^*(z)} ~;
\end{equation}
$M_{\rm Pl}=2.4 \times 10^{18}~\rm GeV$ is the four-dimensional (reduced) Planck scale and $z$ is the cosmic redshift. Notice that the dispersion relation (\ref{stringdr}) is always \emph{sub-luminal} for specifically stringy reasons. The bound $M_{\rm QG} \ge 1.22 \,M_{\rm Pl}$~\cite{GRB2} on the Quantum Gravity (QG) scale can be thus interpreted as an upper bound on the linear density of defects $n^*(z)$, which, in an inhomogeneous D-material universe, depends in general on the redshift.

A potential association of the D-particle defects, which are massive with masses $M_{\rm s}/g_{\rm s}$ (with $M_{\rm s}$ the string scale and $g_{\rm s} < 1$ the weak string coupling), with dark matter has been made in Ref.~\cite{shiu}. A detailed analysis within a concrete microscopic framework showing that, within the framework of the D-material universe, the amount of required conventional dark matter is reduced, whilst in addition the model offers a natural mechanism for the growth of large-scale structure and a successful inflationary scenario, has been developed in Refs.~\cite{emsy}, \cite{msy}.

The main ingredient responsible for the interesting features of the D-material universe model in inducing a large scale structure in the universe~\cite{msy} but also a period of inflation~\cite{emsy} in the absence of an inflaton field with a fine-tuned potential, is the recoil-velocity of the D-particles during their interactions with the stringy matter, which leads to a vector field. Interestingly, the non-linear Born-Infeld type dynamics of the D-matter recoil velocity vector field allows~\cite{msy,odintsov} for the formation of scalar condensates of the corresponding field strength $\llang F_{\mu\nu} \,F^{\mu\nu} \rrang$ which is viewed as a homogeneous scalar field with a mild time-dependence, virtually constant within a given cosmological era. Its value though differs in general from era to era, hence at an inflationary era, due to the dense D-particle populations as assumed in Ref.~\cite{emsy}, the value of the condensate is much larger than the one at late-time eras, like for redshifts $z < 10$, where the astrophysical sources for the observed GW~\cite{LIGO} are located.

In the presence of D-particle ensembles, both the pattern of emission and the propagation of GW will in principle be modified. The modification of the GW emission pattern due to the presence of D-particles in the region of the collapsing black holes may be expected to be negligible in the sense that the ensemble of massive D-particles will behave as matter in the presence of the spiralling black hole system, and the gravitational pull they will exert on the black holes will be very weak to affect the formation of the giant black hole and the subsequent emission of GW.

However, this is not the case for the velocity of propagation of gravitons in the medium, far away from the black hole source, which will be affected in two ways, as we shall discuss in the following. Firstly, the propagation speed of GW will be reduced as compared to the massless case (\emph{sub-luminal} propagation), due to the development of a mass, as a result of the (gravitational) interaction with the recoil-velocity condensate field. Secondly, the presence of dark energy density in the universe, either as a result of the recoil kinetic energy of the D-particles or due to additional dark matter species in the universe (that may co-exist with the D-particles), will also induce a {\it super-luminal} contribution to the group velocity of gravitons. Current observations, including GW interferometry, can provide restrictions to such effects in a way that will be the topic of our discussion here.

The structure of the article is as follows: In Section~\ref{sec:mass}, we discuss the effect of the induced graviton-mass, as a consequence of the graviton propagation in the D-matter ``medium'' with non trivial recoil-velocity condensate fields. In Section~\ref{sec:DM}, we discuss the refractive index effects as a result of the finite energy density of D-particles and other species of dark matter in the universe. In Section~\ref{sec:pheno}, we study the phenomenology of these effects using results from the recent aLIGO GW detection and observations involving ultra-high-energy cosmic rays. Our analysis leads to constraints on the parameters of the model, in particular lowering significantly the maximal allowed magnitude of the string scale itself, under some natural assumptions.

%%%!!!%%%

%%%%%%%%%%%%%%%%%%%%%%%%%%%%%%%%%%%%%%%%%%%%%%%%%%%%%%%%%%
\section{Induced graviton-mass in D-material universe \label{sec:mass}}

One of the most important r\^ole of the D-matter recoil-field condensate lies on its effects on the graviton equation of motion where, along with a modification in the gravitational constant in the string frame description, it contributes to a mass term for the graviton, leading to an additional polarisation mode~\cite{thorn}. We shall discuss this issue next, while later on in the article we shall discuss the implications of the current bounds on the graviton mass in terms of the D-particle density and mass $M_{\rm s}/g_{\rm s}$ that enter the respective formulae.

%%===========================================================================%%
\subsection{Theoretical considerations}

We commence our discussion from the effective (low-energy) action describing the interaction of the vector recoil-velocity field $A_\mu$ with the graviton in the string frame (with respect to the dilaton $\phi$)~\cite{emsy}
\begin{align} \label{BIact}
	S_{\rm eff~4D} =& \int {\rm d}^4 x \left[ - \frac{T_3}{g_{s0}} e^{-\phi} \sqrt{- {\rm det} \left( g + 2\pi \alpha^\prime \,F \right)} \left( 1 - \alpha \,R(g) \right) - \sqrt{-g} \,\frac{1}{4} \langle {\mathcal G}_{\mu\nu} \,{\mathcal G}^{\mu\nu} \rangle \right. \nn
	& \left. - \sqrt{-g} \,e^{-2 \phi} \left( - \frac{1}{\kappa_0^2} \,{\tilde \Lambda} + \frac{1}{\kappa_0^2} \,R(g) + \lambda \left( A_\mu \,A^\mu + {\rm const} \right) + {\mathcal O} \left( \left( \partial \phi \right) \right) \right) \right] ~,
\end{align}
where $\bf{F} = \mathbf{d} \wedge \mathbf{A}$ is the field strength of the vector field and $-\frac{1}{4} \langle {\mathcal G}_{\mu\nu} \,{\mathcal G}^{\mu\nu} \rangle $ is a dilaton-independent term of flux fields, assumed condensed ($\langle \dots \rangle$), which is induced by the bulk geometry and serves the purpose of keeping any cosmological constant terms on the brane universe in the current era (we are interested in here) positive and small, in accordance to observations~\cite{Planck}. We shall define the various other quantities appearing in Eq.~(\ref{BIact}) below.

We consider constant dilaton fields $\phi=\phi_0$ in the galactic era and weak recoil fields $\sqrt{\alpha^\prime} \,A_\mu \ll 1$ (appropriate for late eras of the universe), in which case the above action is well approximated by (in the Einstein frame with respect to the dilaton $\phi$)~\cite{emsy}
\begin{align} \label{effact}
	S^{\rm E}_{\rm eff~4D} = \int {\rm d}^4x &\sqrt{-g} \left[ \left( \frac{1}{2} M_{\rm Pl}^2 + \frac{\alpha e^{-2 \phi_0} \,\tilde F_{\mu\nu} \tilde F^{\mu\nu}}{4} \right) R - 2 \Lambda_0 - \frac{1}{4} \langle {\mathcal G}_{\mu\nu} \,{\mathcal G}^{\mu\nu} \rangle \right. \nn
	& \qquad \left. - \frac{1}{4} \,\tilde F_{\mu\nu} \tilde F^{\mu\nu} + \lambda \left( \tilde A_\mu \tilde A^\mu + \frac{1}{\alpha'} {\cal J} \right) \right] + S_{\rm m}~,
\end{align}
with
\begin{align}
	{\cal J} &\equiv \frac{(2 \pi \alpha')^2 \,T_3 \,e^{3 \phi_0}}{g_{{\rm s}0}}~, \nn
	\frac{1}{2} M_{\rm Pl}^2 &\equiv \frac{1}{16 \pi \,G} = \frac{\alpha \,T_3 \,e^{\phi_0}}{g_{{\rm s}0}} + \frac{1}{\kappa_0^2} ~, \nn
	\Lambda_0 &\equiv \frac{T_3 \,e^{3\phi_0}}{2 g_{{\rm s}0}} + \frac{\tilde{\Lambda} \,e^{2 \phi_0}}{2 \kappa_0^2} ~. \label{Lambda0}
\end{align}
The tilded vector field $\tilde A_\mu$ in the action (\ref{effact}), as compared to the original action (\ref{BIact}), results from an appropriate normalization so that the vector field appears with a canonical (Maxwell) term for its field strength $\tilde F_{\mu\nu}$. The quantity $\lambda$ is a Lagrange multiplier field implementing the constraint on the recoil-velocity field stemming from its interpretation as a velocity four-vector field~\cite{msy}, $M_{\rm Pl} = 2.4 \times 10^{18}~\rm GeV$ is the reduced Planck mass in four space-time dimensions (on our brane universe), $\tilde \Lambda$ is a bulk cosmological constant, which can be constrained (\emph{cf.} discussion below Eq.~(\ref{cosmoconstant})) by means of the dilaton equation of motion, $T_3 > 0$ is the three-brane universe tension, $\alpha^\prime = M_{\rm s}^{-2}$ is the Regge slope, with $M_{\rm s}$ the string scale, and $g_{{\rm s}0}$ the string coupling. For the rest of this work we assume the phenomenological value $g_{{\rm s}0}^2/(4\pi) = 1/20$ that is $g_{{\rm s}0} \sim 0.8$, for which string perturbation theory is valid. Moreover, the reader should recall~\cite{emsy} that, under the assumptions that $\tilde F_{\mu\nu} \,\tilde F^{\mu\nu}$ is almost constant and $\alpha R \ll M_{\rm Pl}^2$, which are appropriate for late eras of the universe we are interested in here, the dilaton equation of motion in the action (\ref{effact}) leads to an expression of $\tilde \Lambda$ in terms of the brane tension $T_3$ which can be used to obtain
\be \label{cosmoconstant}
	\Lambda_0 \simeq - \frac{1}{2} \frac{T_3 \,e^{3 \phi_0}}{g_{{\rm s}0}} < 0 ~.
\ee
This anti-de Sitter type cosmological constant would not be phenomenologically acceptable in the current era, as it would defy the Cosmic Microwave Background (CMB), Baryon Acoustic Oscillation (BAO) and gravitational lensing data. To remedy this fact we assume~\cite{emsy} that contributions from bulk flux gauge fields ${\mathcal G}_{\mu\nu}$ (that condense) and distant (to the brane) D-particles amount to a positive cosmological constant type vacuum energy contribution that fine tunes the negative cosmological constant (\ref{cosmoconstant}) to an acceptably small positive amount in the current era. This assumption will be understood in what follows in the sense that
\be
	\Lambda^{\rm vac} \equiv \Lambda_0 + \frac{1}{8} \langle {\mathcal G}_{\mu\nu} \,{\mathcal G}^{\mu\nu} \rangle + \dots > 0 ~,
\ee
where $\dots$ denote other bulk D-particle contributions to the brane vacuum energy, so that $\Lambda^{\rm vac}$ is compatible with the bounds on the cosmological constant $\Lambda$ from observations in the context of the $\Lambda$CDM (Cold Dark Matter with a positive cosmological constant $\Lambda$) model~\cite{Planck}.

As discussed in detail in Refs.~\cite{mavro, msy,emsy}, the D-particle recoil is in general described by a vector field excitation with two types of contributions:

(i) ``Electric type'', associated with the linear recoil momentum excitations, described by $\sigma$-model world-sheet boundary ($\partial \Sigma$) deformations of the form
\be \label{lm}
	V_{\rm lin.~mom.} = \frac{1}{2\pi \alp} \,\int_{\partial \Sigma} {\rm d} \tau \,g_{ik} \,u^k \,X^0 \,\Theta_\epsilon (X^0) \,\partial_n X^i ~,
\ee 
where $\Theta_\epsilon (X^0)$ is a regularised Heaviside function, describing the impact of the matter string on the D-particle at a time $X^0=0$ and $\partial_n X^i$ denotes the normal world-sheet derivative. The quantity $u^i$ is the D-particle recoil-velocity and $g_{ij}=a^2(t) \delta_{ij}$ are the spatial components of the metric for a (spatially flat) Friedmann-Lema\^{i}tre-Robertson-Walker (FLRW) universe, with scale factor $a(t)$, we assume here, as dictated by the current astrophysical/cosmological data~\cite{Planck}. For the galactic eras we are interested in this work, one can take $a(t) \sim 1$, which will be assumed from now on. The vertex operators (\ref{lm}) satisfy a (logarithmic) conformal algebra on the world-sheet~\cite{kogan}, hence they are consistent string deformations. They correspond to vector field excitations $\tilde A_i$ with a target-space-time field strength (after the impact) of the form
\be \label{electric}
	 \tilde F_{0i} = E_i = M_{\rm s}^2 \,g_{ij} u^j ~,
\ee
where $E_i $ denotes the ``electric'' field.

(ii) ``Magnetic type'', associated with $\sigma$-model deformations corresponding to non-zero angular momentum of the recoiling D-particles, described by the world-sheet boundary vertex operators~\cite{mavro}
\be \label{am}
	V_{\rm ang.~mom.} = \frac{1}{2\pi \alpha^\prime} \,\int_{\partial \Sigma} {\rm d} \tau \,\epsilon_{ijk} \,u^k \,X^j \Theta_\epsilon (X^0) \,\partial_n X^i ~,
\ee
with $\epsilon_{ijk}$ the antisymmetric symbol in three-space dimensions. As is the case of the ``impulse'' linear momentum vertex operators (see (\ref{lm}))~\cite{kogan}, the operators (\ref{am}) also satisfy~\cite{mavro} a (logarithmic) conformal algebra on the world sheet of the string. These imply a target-space field strength with spatial components
\be \label{magnetic}
	F^{ij} = - \epsilon^{ijk} \,B_k = M_{\rm s}^2 \epsilon^{ijk} g_{k\ell} \,u^\ell \,\Rightarrow \,B_k = M_{\rm s}^2 \,g_{k\ell} \,u^\ell ~,
\ee
where $B_i$ denotes the ``magnetic'' field.

Although in the gravitational lensing analysis~\cite{emsy} we have ignored the angular momentum contributions, which as we shall see do not change the order of magnitude of our conclusions, nevertheless for the purpose of our present analysis, which is to study gravitational wave propagation in the D-material universe in the (low-temperature, compared to the inflationary epoch) galactic era, such contributions shall play an important r\^ole for the stability of the vacuum. For the (unstable) inflationary high-temperature phase, such contributions are negligible~\cite{emsy} and thus the conclusions of our previous work remain unchanged.
 
The graviton equation of motion obtained from the action (\ref{effact}) reads
\be \label{EOMgraviton}
	\left( R_{\mu\nu} - \frac{g_{\mu\nu}}{2} R \right) \left[ \frac{1}{2} M_{\rm Pl}^2 + \frac{\alpha e^{-2\phi_0} \,\tilde F^2}{4} \right] = \frac{1}{2} \,T_{\mu\nu}^{\rm rec} - \,g_{\mu\nu} \,\Lambda^{\rm vac} + \frac{1}{2} \,T_{\mu\nu}^{\rm m}~, 
\ee
where from now on we use the short-hand notation $\tilde F^2 = \tilde F_{\mu\nu} \tilde F^{\mu\nu}$. Note that $T^{\rm m}_{\mu\nu}$ denotes the stress tensor of conventional matter, including dark matter other than D-particles, and $T_{\mu\nu}^{\rm rec} $ is the recoil-velocity contribution
\be \label{stress}
	T_{\mu\nu}^{\rm rec} = \tilde F_{\mu \alpha} \,\tilde F_{\nu}^{ \,\,\alpha} - g_{\mu\nu} \,\frac{\tilde F^2}{4} ~.
\ee
The latter resembles of course the corresponding stress tensor of electrodynamics, but here the vector field $\tilde A_\mu$ is the recoil-velocity field, which satisfies the constraint\footnote{This is the only effect of the Lagrange multiplier field $\lambda$. Indeed, as the analysis of Ref.~\cite{msy,emsy} has demonstrated, any terms in the equations of motion involving the field $\lambda$ become --- upon its expression, via the equations of motion, in terms of the other fields in the Lagrangian --- proportional to terms with gravitational-covariant derivatives acting on $\tilde F$, which are negligible under our assumptions here.}~\cite{msy,emsy}
\be
	\tilde A_\alpha \tilde A^\alpha + \frac{1}{\alpha'} {\cal J} = 0 ~. \nn
\ee
A few remarks are in order here. The dynamics of the vector recoil field $\tilde A_\mu$ in the action (\ref{effact}) is much more complicated than the lowest-order weak-field expansion given above. Actually, as discussed in Ref.~\cite{msy}, detailed string theory considerations imply that there is a Born-Infeld term, whose perturbative expansion yields the Maxwell kinetic term in the action (\ref{effact}). Such non-linear square root interactions may be responsible for the formation of condensates of the recoil-velocity field, following the discussion in Ref.~\cite{odintsov}, which was adapted to the D-matter case in Refs.~\cite{msy,emsy}. Therefore, from now on we assume that $\tilde F^2$ can condense, forming a scalar-like field, which is at most time-dependent at cosmological scales. We thus have
\be \label{cond}
	\sigma_F (t) \equiv \langle \tilde F^2 \rangle = \llang \tilde F^2 \rrang + \langle \tilde F^2 \rangle_{\rm q} ~,
\ee
where $\llang \dots \rrang $ denote {\em classical} condensates, due to the statistical nature of the recoil velocity field in macroscopic D-particle populations in the universe, whose magnitude has been estimated in Ref.~\cite{emsy}, while $\langle \dots \rangle_{\rm q}$ denotes \emph{quantum vacuum} effects~\cite{odintsov}, associated with the full Born-Infeld dynamics of the vector field, which cannot be computed at present. Since our point here is to study GW propagation from sources at redshifts $z < 10 $, as is the situation characterising the recent discovery reported in Ref.~\cite{LIGO}, where $z \sim 0.09$, we consider short enough scales for which $\sigma_{\rm F}$ is practically constant, thus suppressing all its derivatives. Of course between cosmological eras the value of $\sigma_F$ changes, in particular at the inflationary era, where strong condensates of the field $\sigma_F$ are needed to drive inflation. For the matter-dominated era, of interest to us here, $\sigma_F $ can be safely assumed to be weak.

In a mean-field approximation, one may first consider (\ref{EOMgraviton}) with the stress tensor of the recoil field averaged in the sense of (\ref{cond}). If we consider \emph{equal strength} electric and magnetic contributions, given respectively by (\ref{electric}) and (\ref{magnetic}), then we get
\be \label{totalcond}
	\sigma_F = \langle \tilde F^2 \rangle = 2 \,\langle \tilde F_{0i} \,\tilde F_{0j} \rangle \,g^{00} g^{jk} + \langle \tilde F_{ik} \,\tilde F_{j\ell} \rangle \,g^{ij} g^{k\ell} ~.
\ee
For the classical statistical averages, we have
\begin{align} \label{classcond}
	\llang \tilde F_{0i } \,\tilde F_{0j} \,g^{ij} \rrang &= M_{\rm s}^4 \,\llang u^i u^j g_{ij} \rrang > 0 ~, \nn
	\llang \tilde F_{0i } \,\tilde F_{0j} \,g^{00} g^{ij} \rrang &= - M_{\rm s}^4 \,\llang u^i u^j g_{ij} \rrang < 0 ~, \nn
	\llang \tilde F_{ik} \,\tilde F_{j\ell} \,g^{ij} g^{k\ell} \rrang &= 2 M_{\rm s}^4 \,\llang u^i u^j g_{ij} \rrang > 0 ~,
\end{align}
and hence, on account of (\ref{totalcond}), we recover the equipartition theorem for the classical condensates of the vector field we are familiar with from ordinary electrodynamics, according to which the classical condensate {\em vanishes}, namely
\be \label{zero}
	\llang \tilde F^2 \rrang = 0 ~.
\ee
We thus have for the appropriately averaged recoil stress tensor (\ref{stress})
\be \label{stress2}
	\llang T^{\rm rec}_{\mu\nu} \rrang = \llang \tilde F_{\mu \alpha } \,\tilde F_{\nu}^{~\alpha} \rrang - \frac{1}{4} \,g_{\mu\nu} \,\llang {\tilde F}^2 \rrang~, 
\ee
which, on account of eqs.~(\ref{classcond}), (\ref{zero}), leads to
\be \label{rhomdmatter}
	\llang T^{\rm rec}_{00} \rrang \equiv \rho_{\rm rec}^{\rm class} = \frac{1}{2} \,\llang E_iE^i \rrang + \frac{1}{2} \,\llang B_i B^i \rrang = \frac{M_{\rm s}^4}{a^2(t)} \,\llang u^i u^j \delta_{ij} \rrang ~, 
\ee
The reader should notice that $\rho_{\rm rec}^{\rm class}$ is of the same order of magnitude as the recoil energy density considered in Ref.~\cite{emsy}, where only ``electric'' type $E_i$ fields were considered (the result is larger by a factor of 2) and hence the gravitational lensing phenomenology conclusions of Ref.~\cite{emsy} remain unchanged.

The quantum fluctuations of the recoil-velocity field are significant in the low-temperature galactic eras and for those we have, as dictated by the isometry structure of the FLRW cosmological space-time~\cite{odintsov}
\begin{align} \label{qcond}
	{\langle \tilde F_{0\alpha} \,\tilde F_0^{\,\,\alpha}\rangle}_{\rm q} &= \frac{\tilde a_t(t)}{4} \,g_{00} ~, \nn
	{\langle \tilde F_{i\alpha} \,\tilde F_j^{\,\,\alpha}\rangle}_{\rm q} &= \frac{\tilde a_{\rm s}(t)}{4} \,g_{ij} ~, \nn
	\sigma_F &= \tilde a = \langle \tilde F^2 \rangle_{\rm q} = \frac{1}{4} \left( \tilde a_{\rm t} + 3 \,\tilde a_{\rm s} \right) > 0 ~,
\end{align}
where $\tilde a_{\rm t}=\tilde a_{\rm t}(t)$ and $\tilde a_{\rm s}=\tilde a_{\rm s}(t)$. Note that we assume the positivity of the quantum condensate $\tilde a$, so as to be able to use such condensates as providers of zero-point (vacuum) energy of de Sitter type~\cite{odintsov,msy}. The corresponding contribution to the recoil stress tensor is then
\begin{align} \label{qstress}
	\langle T^{\rm rec} _{00} \rangle_{\rm q} &= - \frac{1}{4} \left( \tilde a - \tilde a_{\rm t} \right) g_{00}~, \nn
	\langle T^{\rm rec} _{ij} \rangle_{\rm q} &= \frac{1}{12} \left( \tilde a - \tilde a_{\rm t} \right) g_{ij} ~.
\end{align}
Another important point we wish to make is that in the current work we view any vacuum energy contribution, including those obtained from the bulk dynamics, as microscopic, due to the (quantum) dynamics of fields of the underlying string theory, and hence related to the stress tensor (right-hand-side of the (low energy) Einstein equations (\ref{EOMgraviton})), rather than geometric in origin thereby related to the left-hand-side. In the latter case one would have  to deal with (anti)de Sitter space times, since those are the maximally symmetric space-times about which 
one expands, in which case the concepts of the graviton mass and the refractive index, upon which we shall concentrate here, become more complicated. For our purposes in the current article we take the point of view that there should be always a flat limit of the left-hand-side of the Einstein's equations, since the result of any cosmological-constant-type term is due to some sort of condensate (either bulk field or recoil D-particle fluctuations). This allows for a conventional definition of GW and massive-graviton effects in the GW propagation, which will be the focus of our attention in what follows. 

With the above in mind, one can then expand the metric around its (non-flat) unperturbed cosmological value $g_{\mu\nu} = g^{\rm (0)}_{\mu\nu} + h_{\mu\nu}$, where the background $g_{\mu\nu}^{(0)}$ takes into account the presence of a (field-induced) ``cosmological-constant-type'' vacuum energy and $|h_{\mu\nu}| \ll 1$. Working, as appropriate for GW analysis, in the transverse traceless (TT) gauge~\cite{thorn}, for which
\be \label{TTgauge}
	\partial_\mu h^\mu_{~\nu} = 0 ~, \quad h^\alpha_{~\alpha} = 0 ~, \quad h_{\mu 0} = 0 ~,
\ee
the perturbed Einstein tensor becomes\footnote{Our conventions are $(-,+,+,+)$ for the signature of the metric, and $R_{\mu\nu} \equiv R^\alpha_{\;\,\mu\nu\alpha} \equiv \partial_\alpha \Gamma^\alpha_{\nu\mu} - \partial_\nu \Gamma^\alpha_{\alpha\mu} + \Gamma^\alpha_{\alpha\beta} \Gamma^\beta_{\nu\mu} - \Gamma^\alpha_{\nu\beta} \Gamma^\beta_{\alpha\mu}$.}
\be
	R_{\mu\nu} - \frac{1}{2} \,g_{\mu\nu} R = - \frac{1}{2} \,\partial^2 h_{\mu\nu} ~.
\ee
In the TT gauge, the only non-zero contributions to the recoil stress tensor to first order in the metric expansion (indicated by the superscript (1)) are the spatial ones
\begin{align} \label{ttstress}
	\llang \tilde F_{i \alpha} \,\tilde F^{\,\alpha}_j \rrang^{(1)} &= \llang \tilde F_{i k} \,\tilde F_{j\ell} \,h^{k\ell} \rrang = M_{\rm s}^4 \,\epsilon_{ikm} \epsilon_{j\ell n} \,\llang u^m \,u^n \rrang \,h^{k\ell} \nn
	&= \delta_{j[i} \,\delta_{k]\ell} \,\frac{1}{3} \,\sigma_0^2 \,h^{k\ell} = -\frac{1}{3} \,\sigma_0^2 \,h_{ij}~,
\end{align}
where we used $h_{ij} = h_{ji}$ and $\sigma_0^2 = M_{\rm s}^4 \,\llang u^m u^n g^{(0)}_{mn} \rrang$, as well as~\cite{msy}
\be
	M_{\rm s}^4 \,\llang u_m u_n \rrang = \frac{1}{3} \,\sigma_0^2 \,g_{mn}^{(0)} \simeq \frac{1}{3} \,\sigma_0^2 \,\delta_{mn} ~, \nn
\ee
since for the galactic era $g_{mn}^{(0)} = a^{2}(t) \,\delta_{mn} \simeq \delta_{mn}$.

Recalling that the zeroth order (in the metric expansion) equation of motion is satisfied and taking into account Eqs.~(\ref{stress}), (\ref{classcond}), (\ref{zero}), (\ref{qcond}) and (\ref{ttstress}), one obtains a first-order equation of motion for the spatial perturbations $h_{ij}$ in the TT gauge (\ref{TTgauge}) of the form
\begin{align} \label{masseq}
	\partial^2 h_{ij} &- \kappa_{\rm eff}^2 \left[ \frac{1}{3} \,M_{\rm s}^4 \,\llang u^m u^n \,g_{mn}^{(0)} \rrang - \frac{1}{12} \left( \tilde a - \tilde a_{\rm t} \right) + 2 \Lambda^{\rm vac} \right] h_{ij} = 0 ~, \nn
	\mbox{where~} \ & \frac{1}{\kappa_{\rm eff}^2} \equiv \frac{1}{\kappa_0^2} + \frac{\alpha \,T_3 \,e^{\phi_0}}{g_{s0}} + \frac{\alpha e^{-2\phi_0} \,\sigma_{\rm F}}{4} ~.
\end{align}
Assuming that the condensate $\sigma_F$ is small and that $\Lambda^{\rm vac}$ is also small as compared to $M_{\rm Pl}^4$, then to leading order in $\sigma_F$ and $\Lambda^{\rm vac}$, one may replace from now on $\kappa_{\rm eff}^2$ by $2 \,M_{\rm Pl}^{-2}$. Hence, Eq.~(\ref{masseq}) is just the equation of motion of a \emph{massive} graviton, with mass squared
\be \label{masscond1}
	m_{\rm G}^2 (t) \simeq M_{\rm Pl}^{-2} \left[ \frac{2}{3} \,M_s^4 \,\llang u^m u^n \delta_{mn} \rrang - \frac{1}{6} \left( \tilde a - \tilde a_{\rm t} \right) + 2 \rho_{\Lambda^{\rm vac}} \right] ~,
\ee
where $\rho_{\Lambda^{\rm vac}} \equiv 2 \,\Lambda^{\rm vac}$.\footnote{Note indeed that our $\Lambda^{\rm vac}$ is of dimension of an energy density, as one can see in the action~(\ref{effact}) or in the definition given below in Eq.~(\ref{Lambda0}).}

The mass is real, provided the right-hand-side of Eq.~(\ref{masscond1}) is positive, otherwise the graviton would appear \emph{tachyonic} and the stability of the vacuum, let alone causality~\cite{shore} due to the \emph{super-luminality} at all frequencies of the corresponding group velocity, would be in jeopardy. Fortunately, this can be easily guaranteed by assuming either small quantum corrections compared to the statistical classical terms or that the condensates $\tilde a$ and $\tilde a_{\rm t}$ are both positive. The latter assumption is in line with attempts~\cite{odintsov}, in the context of Born-Infeld electrodynamics, to associate such quantum condensates with positive (de Sitter type) contributions to the vacuum energy. We shall thus make this assumption in what follows.

In this latter respect, from Eqs.~(\ref{classcond}) and (\ref{qstress}), we observe that the recoil energy density, including quantum condensate contributions, reads
\be \label{red}
	\rho_{\rm rec}^{\rm full} = M_{\rm s}^4 \,\llang u_i u_j \delta^{ij} \rrang + \frac{\tilde a - \tilde a_{\rm t}}{4} > 0 ~.
\ee
We now impose the requirement that the upper bound of $\rho_{\rm rec}^{\rm full}$ should {\em not} exceed the matter energy density $\rho_{\rm m}^{\Lambda {\rm CDM}}$ of the $\Lambda$CDM model. For the value of $\rho_{\rm m}^{\Lambda {\rm CDM}}$ we take here the benchmark point~\cite{Planck}
\be \label{bench}
	\rho_{\rm m}^{\Lambda \rm CDM} = 0.3 \,\rho_{\rm c}^0 = 0.9 \,H_0^2 \,M_{\rm Pl}^2 = 9 \times 10^{-121}~M_{\rm Pl}^4 ~,
\ee
with $\rho_{\rm c}^0$ the current-era critical density and $H_0 \sim 10^{-60} \,M_{\rm Pl}$ the present-day Hubble rate. Hence we obtain
\be \label{red4a}
	0 < \rho_{\rm m} \equiv \rho_{\rm rec}^{\rm full} + \rho_{\rm DM+b} \sim \rho_{\rm m}^{\Lambda \rm CDM} ~,
\ee
and
\be \label{red4b}
	\frac{M_{\rm s}^4}{a^2(t)} \,\llang u_i u_j \delta^{ij} \rrang + \frac{\tilde a - \tilde a_{\rm t}}{4} \lesssim \rho_{\rm m}^{\Lambda \rm CDM} ~,
\ee
where $\rho_{\rm m}$ is the total matter energy density of the universe, including D-matter as well as ``conventional'' dark matter and baryonic matter (denoted together as $\rho_{\rm DM+b}$) contributions, which, according to Ref.~\cite{emsy}, would imply that the recoil-velocity contributions in the D-material universe would be compatible with the $\Lambda$CDM model.

If the upper bound in the inequality of (\ref{red4b}) is saturated, then D-matter provides the dominant component of dark matter~\cite{shiu}. The reader should recall though that the Born-Infeld form of the recoil velocity vector field $\tilde A_\mu$ studied here and in Refs.~\cite{msy,emsy} provide a dark fluid which also contributes to dark energy, hence recoiling D-matter should be viewed as a {\em mixed} dark-energy/dark-matter model.

In this respect, the condition (\ref{red4a}) also ensures that the total energy density of the D-material universe, including vacuum energy contributions
\be \label{rhototal}
	\rho_{\rm total} = \rho_{\rm m} + \rho_{\Lambda^{\rm vac}} ~,
\ee
is of the order dictated by the current data~\cite{Planck}, i.e. close to the critical density. Thus, the conclusions of Ref.~\cite{emsy} that D-matter can play the r\^ole of dark matter in galactic lensing measurements are still valid, given that the order of magnitude of the contributions to the recoil energy density did not change by the inclusion of ``magnetic'' field (\ref{magnetic}) components in the Born-Infeld fluid describing the recoil excitations of the D-particles.

Let us make a short remark on the order of magnitude of the allowed density of D-particles in the D-material universe~\cite{msy}. We recall that in the galactic era, one has the following estimate for the statistical (classical) component of the recoil-velocity condensate~\cite{emsy}
\be \label{u2}
	\llang u_i u_j \delta^{ij} \rrang \sim \frac{N_{\rm D}^0}{N_\gamma^0} \frac{{\widetilde \xi}_0^2 \,|\vec{p}_{\rm phys}|^2}{M_{\rm s}^2 } \,g_{{\rm s}0} ^2 ~,
\ee
with $\tilde \xi_0 < 1$ an order ${\cal O}(1)$ parameter, that describes the momentum transfer during the scattering of a D-particle with an open string representing radiation (which is assumed to be the dominant species with which the D-particles interact). The quantity ${\vec p}_{\rm phys}$ is the ``physical'' average $3$-momentum of a photon as observed by a comoving cosmological observer in the FLRW universe, assumed to be a thermalised CMB photon at $T = 2.7~{\rm K}$, hence $|\vec{p}_{\rm phys}| \simeq 3 k_{\rm B} \, T \simeq 7.2 \times 10^{-4}~{\rm eV} \simeq 3 \times 10^{-31}~M_{\rm Pl}$. By $N_{\rm D}^0$ and $N_\gamma^0$ we denote the number densities of D-particles and photons, respectively, in the current era of the universe; note that $N_\gamma^0 = 4 \times 10^{-97}~M_{\rm Pl}^3$~\cite{Planck}. In deriving (\ref{u2}) we assumed $N^0_{\gamma} \gg N^0_{\rm D}$, so that $\sfrac{N^0_{\rm D}}{(N^0_{\gamma} + N^0_{\rm D})} \simeq \sfrac{N^0_{\rm D}}{N^0_{\gamma}}$ is the probability of interaction of D-particles with the CMB photons that constitute the most dominant species for the recoil of D-particles in the medium.

We also note that the analysis of Ref.~\cite{msy,emsy} implied a lower limit to the density of D-particles, as a result of the requirement that the D-matter can enhance the growth of large-scale structure in the universe. In fact, if we ignore (assuming them as sub-leading) the quantum corrections in Eq.~(\ref{red}), then, in view of the inequality (\ref{red4b}), we get the following bounds on the statistical condensate $\llang u_i u_j \delta^{ij} \rrang$ defined in (\ref{u2})
\be \label{Nrange}
	10^{-123} \,\frac{M_{\rm Pl}^2}{M_{\rm s}^2} \lesssim \llang u_i u_j \,\delta^{ij} \rrang \lesssim 10^{-120} \,\frac{M_{\rm Pl}^2}{M_{\rm s}^2} ~,
\ee
which lead to the following bounds on the D-particle density $N_{\rm D}^0$
\be
	10^{-123} \,\frac{M_{\rm Pl}^2}{g_{{\rm s}0}^2 \,{\widetilde \xi}_0^2 \,|\vec{p}_{\rm phys}|^2} \lesssim \frac{N^0_{\rm D}}{N_\gamma^0} \lesssim 10^{-120} \,\frac{M_{\rm Pl}^2}{g_{s0}^2 \,{\widetilde \xi}_0^2 \,|\vec{p}_{\rm phys}|^2} ~,
\ee
which turn out to be independent of $M_{\rm s}$:
\be \label{NrangeNum}
	6 \times 10^{-159} \,{\widetilde \xi}_0^{-2} \,M_{\rm Pl}^3 \;\lesssim N_{\rm D}^0 \lesssim \; 6 \times 10^{-156} \,{\widetilde \xi}_0^{-2} \,M_{\rm Pl}^3 ~.
\ee
These estimates are affected if the quantum fluctuations $\tilde a$, $\tilde a_t$ to the condensate $\sigma_F$ are included. Unfortunately, lacking a microscopic theory of stringy D-particles we cannot estimate the magnitude of the quantum condensates $\tilde a$, $\tilde a_t$ entering the mass (\ref{masscond1}) and hence we can only discuss below some phenomenological bounds coming from experimental constraints on the graviton mass. At any rate for the galactic eras of relevance to us today we assume that the quantum fluctuations are of the same order as the statistical condensate.

%%===========================================================================%%
\subsection{Phenomenological constraints on induced graviton-mass and implications for the D-material universe} 

To discuss effects of matter in the GW propagation, let us first remark that the relativistic dispersion formula for {\em massive} gravitons $\omega^2 = k^2 + m_{\rm G}^2$ (in natural units), leads to the sub-luminal group velocity (denoted by a subscript ${\rm g}$)
\be \label{gvelg}
	v_{\rm g}^{\rm mass} = \frac{\partial \omega}{\partial k} = \frac{k}{\omega} = \frac{1}{v_{\rm p}^{\rm mass}} = n^{\rm mass}_{\rm G} \simeq 1 - \frac{m_{\rm G}^2}{2 \,\omega^2}~,
\ee
assuming $m_{\rm G} \ll \omega$, where $n^{\rm mass}_{\rm G}$ denotes the index of refraction of GW due to the graviton mass and $v_{\rm p}^{\rm mass}$ is the corresponding phase velocity (which is larger than unity, without conflict with causality, as the front of the wave does not carry out any physical information). For two gravitons with frequencies $\omega$ and $\omega'$, the difference in group velocities is thus
\be \label{gvelg2}
	\Delta v_{\rm g}^{\rm mass} = \frac{m_{\rm G}^2}{2} \left| \frac{1}{\omega^2} - \frac{1}{\omega'^2 } \right| ~.
\ee
The induced dispersion in the GW, taking into account the cosmic expansion (redshift $z$) of a standard $\Lambda$CDM universe, leads to differences in the observation times of GW components of two different (low) frequencies $\omega$ and $\omega'$, emitted with a time difference $\Delta t_{\rm e}$ at the source~\cite{Will}
\begin{align} \label{dtmass}
	\Delta t^{\rm mass}_{\rm o} &= (1 + z) \left[ \Delta t_{\rm e} + (1 + z) \,{\mathcal D} \,\frac{m_{\rm G}^2}{2} \left( \frac{1}{\omega^2} - \frac{1}{\omega'^2} \right) \right] ~, \nn
	\mbox{where}\ \ {\mathcal D} &= \int_{z_{\rm o}}^{z_{\rm e}} \frac{(1 + \tilde z)^{-2} \,{\rm d} {\tilde z}}{H_0 \sqrt{\Omega_{\rm m} (1 + \tilde z)^3 + \Omega_\Lambda}} ~,
\end{align}
with $D \equiv (1 + z) \,{\mathcal D} = (1 + z) \int_{t_{\rm e}}^{t_{\rm o}} a(t) \,{\rm d}t$ the proper distance, $a(t)$ the scale factor (in units where today $a_0 = a(t_{\rm o}) =1$) and where the subscipt $o$ ($e$) pertains to observation (emission) quantities. In the standard $\Lambda$CDM fiducial cosmology~\cite{Planck}, which we assume here, we have $(\Omega_{\rm m}, \Omega_\Lambda, \Omega_k) = (0.3, 0.7, 0)$, which will be used throughout this work.

Assuming for simplicity that the two gravitons where emitted simultaneously ($\Delta t_{\rm e} \simeq 0$) one may get from (\ref{dtmass}) a lower bound for the graviton mass to be {\em detectable} by interferometric GW devices with time-difference sensitivity $\Delta t_{\rm s}$ and $\omega' = \xi \,\omega$, given by
\be \label{sens}
	m_{\rm G}^2 \,\ge \,\frac{\xi^2}{|1 - \xi^2|} \,\frac{2 \,\Delta t_{\rm s} \,\omega^2 }{(1 + z)^2 \,{\mathcal D}} ~.
\ee
The aLIGO measurements~\cite{LIGO} achieve a very good time-frequency coverage for a broad range of signal morphologies by having the analysis repeated with seven frequency resolutions ranging from 1 Hz to 64 Hz in steps of powers of two, corresponding to time resolutions 
\be\label{timeres}
\Delta t_{\rm s}^{\rm aLIGO} = \sfrac{1}{2} \,(\Delta \omega_{\rm s}^{\rm aLIGO})^{-1} \in [7.8 \times 10^{-3}, \,5 \times 10^{-1}]~{\rm s}~. 
\ee
The clusters at different resolutions overlapping in time and frequency are then combined into a trigger that provides a multi-resolution representation of the excess power event recorded by the detectors. The minimum of the right-hand-side of the inequality (\ref{sens}) is obtained for the minimum value of the time resolution possible, that is in our case $\Delta t_{\rm s}^{\rm aLIGO} \sim 7.8 \times 10^{-3}~{\rm s}$ and the minimum value of $\xi$. Theoretically, if $\Delta \omega =0 $   could be measured experimentally, then the experiment would have infinite sensitivity to measure the graviton mass; however the minimum possible detectable frequency difference is the frequency resolution given by Eq.~(\ref{timeres}), which for the lower limit on $\Delta t_{\rm s}$ considered, leads to $\Delta \omega_{\rm s}^{\rm aLIGO} \simeq 64$~Hz.
With these values, for gravitons in the aLIGO frequency detection range $\omega \simeq 100~{\rm Hz} \simeq 4 \times 10^{-13}~{\rm eV} \simeq 1.7 \times 10^{-40}~M_{\rm Pl}$ emitted at a distance of $410$ Mpc (corresponding to a redshift $z \simeq 0.09$ and hence ${\mathcal D} = 0.08 \,H_0^{-1}$~\cite{LIGO}, with $H_0 \simeq 10^{-60} \,M_{\rm Pl}$), we get
\be \label{lowermg}
	m_{\rm G} \gtrsim 4.6 \times 10^{-50}~M_{\rm Pl} \simeq 1.1 \times 10^{-22}~{\rm eV} ~,
\ee
in order for the graviton mass to be observable by aLIGO. If the time and frequency resolution improves in future interferometric networks, leading to improvements of the signal to noise ratio $\omega \,\Delta t $ smaller than 1/10, value which characterises aLIGO~\cite{LIGO}, then the sensitivity to the graviton mass will increase.

Assuming a standard $\Lambda$CDM cosmology, the LIGO collaboration performed a detailed statistical analysis~\cite{LIGO} during the observation of GW by the black-hole merger event GW 150914, and found no significant signal up to Compton wavelengths of order $\lambda_{\rm q}^{\rm aLIGO} = h/m_{\rm G}^{\rm aLIGO} > 10^{13}~{\rm km}$ , implying an upper bound on the graviton mass
\be \label{ligobound}
	m_{\rm G}^{\rm aLIGO} < 1.2 \times 10^{-22}~{\rm eV} \simeq 5.0 \times 10^{-50}~M_{\rm Pl} \quad {\rm (aLIGO)}~,
\ee
which is in perfect agreement with the analytical bound (\ref{lowermg}). It can be used in our model to bound the condensate effects responsible for the induced graviton-mass (\ref{masscond1}). 

Before doing so, let us discuss first some additional effects of the D-particle ``medium'' on the propagation of GW in the D-material universe. As we shall argue in the next section, D-matter may induce a refractive index for graviton propagation, which leads to additional constraints, beyond the ones discussed due to the induced graviton-mass.

%%%%%%%%%%%%%%%%%%%%%%%%%%%%%%%%%%%%%%%%%%%%%%%%%%%%%%%%%%
\section{Other effects on graviton propagation in the D-material universe \label{sec:DM}}

In addition to the mass induced effects, graviton propagation in the D-material universe (which includes also conventional dark matter components) is also affected by refractive index effects in the medium of D-particles. Given the low-frequency regime ($\omega \sim 100~{\rm Hz}$) of GW of relevance to the LIGO observations, we expect (and shall verify explicitly below) that any stringy effect of the D-foam on the GW propagation --- in general expected to increase with frequency, being proportional to some positive power of it --- is negligible. This leaves the low-energy point-like field theory interactions of GW with the environment of matter (including dark matter) scatterers in the universe as the dominant source of induced refraction for low-frequencies.

%%===========================================================================%%
\subsection{Refractive index of gravitons}

If gravitational waves propagate in a medium of matter scatterers with density $\rho_{\rm m}$, then they will experience an induced refractive index, arising from the coordinate dependent gravitational potential corrections to the Newtonian metric, as demonstrated long ago in Ref.~\cite{Peters}. To estimate such effects it suffices to consider the approximate situation in which all matter is assumed concentrated in a ``thin'' spatial layer of thickness $\Delta z$, through which GW pass. Such layers modify the gravitational Newtonian potential felt by GW. To lowest order in $\omega$, for {\it massless} gravitons, the index of refraction is larger than unity for $\rho_{\rm m} > 0$ and of the form
\begin{equation} \label{refraction}
	0 < n_{\rm G}^{\rm DM} -1 \simeq \frac{2\pi \,G \,\rho_{\rm m}}{\omega^2} = \frac{\rho_{\rm m}}{4\,M_{\rm Pl}^2 \,\omega^2} \ll 1 ~,
\end{equation}
to linear order in the gravitational potential induced by matter. Here, $\rho_{\rm m}$ is the (4-dimensional) matter density (including dark matter and D-matter) (see Eq.~(\ref{bench})); we took that the (4-dimensional) gravitational constant is $8\pi\,G =M_{\rm Pl}^{-2}$ and the frequency range which we are interested in is $\omega \simeq 100$~Hz. In our case, the recoil contribution of the D-material universe is included in $\rho_{\rm m}$ which is then expressed as in Eq.~(\ref{red4a}).

Equatiion~(\ref{refraction}) implies that the phase velocity of gravitational waves, $v_{\rm p}^{\rm DM} = \sfrac{1}{n}$, is {\em sub-luminal} while the group velocity, $v_{\rm g}^{\rm DM}$, is {\em super-luminal} for low-$\omega$. Indeed, to obtain the latter, one can use the derivative of the refractive index with respect to $\omega$:
\be
	\frac{1}{v_{\rm g}} = n + \omega \,\frac{{\rm d} n}{{\rm d} \omega} ~,
\ee
which, in the case of a medium with refractive index given by $n - 1 = \chi \omega^{-2}$ with $\chi$ a constant (as we have here), leads to
\begin{align}
	\frac{1}{v_{\rm g}} = n - 2 \,\chi \omega^{-2} = 1 - \chi \,\omega^{-2} \quad \Rightarrow \quad v_{\rm g} \simeq 1 + \chi \omega^{-2} > 1 ~,
\end{align}
if $\chi \omega^{-2} \ll 1$. Hence, the {\em super-luminal} group velocity for {\em massless} gravitons propagating in the dark matter and D-matter medium, yields here
\be \label{gvelsl}
	v_{\rm g}^{\rm DM} \simeq 1 + \frac{\rho_{\rm m}}{4\,M_{\rm Pl}^2 \,\omega^2} \simeq 1 + 10^{-41} ~,
\ee
where we considered again $\omega \simeq 1.7 \times 10^{-40}~M_{\rm Pl}$.

This will lead to time differences in the arrival times of two gravitons with frequencies $\omega$ and $\omega'$, using Eq.~(\ref{dtmass})) and replacing the term $\sfrac{1}{2} \,m_{\rm G}^2$ by $\sfrac{\rho_{\rm m}}{4 \,M_{\rm Pl}^2}$, yielding
\be \label{dtDM}
	\Delta t^{\rm DM}_{\rm o}= - (1 + z) \left[ \Delta t_{\rm e} + (1 + z) \,{\mathcal D} \,\frac{\rho_{\rm m}}{4 \,M_{\rm Pl}^2} \left( \frac{1}{\omega^2} - \frac{1}{\omega'^2} \right) \right] ~,
\ee
with $(1 + z) \,{\mathcal D}$ the proper distance from the GW source to the observer as in Eq.~(\ref{dtmass}). Note that the relative minus sign in Eq.~(\ref{dtDM}), as compared to Eq.~(\ref{dtmass}), is due to the fact that $\Delta t^{\rm DM}_{\rm o}$ now denotes an advance rather than a delay due to the {\rm super-luminal} nature of the graviton group velocity.

Some comments are in order here regarding the super-luminal nature of the group velocity (\ref{gvelsl}). This was to be expected by the corresponding case for light propagation in a non-trivial vacuum~\cite{latorre}. The graviton excitations find themselves in a {\em negative} (as compared to the trivial flat space-time empty vacuum) gravitational energy density $\rho = - \rho_{\rm m} < 0$ environment (as a result of the attractive gravitational potential of the scatterers exerted on the graviton ``particles''). Indeed, in such non-trivial vacua with an energy density $\rho$, the group velocity of massless photons or gravitons, after taking into account vacuum polarization effects, deviates from 1 by an amount $v_{\rm g} - 1 \propto - \rho = + \rho_{\rm m} > 0$. The (low-frequency) super-luminal GW velocity is not in conflict with causality according to the argumentation in Ref.~\cite{latorre}, since no physical (i.e. observer independent) information can be transmitted, given that the results pertain to a specific frame (Robertson Walker); moreover, it is the high-frequency limit that would be of relevance~\cite{shore}.

%%===========================================================================%%
\subsection{Refractive index of photons}

It should be remarked at this point that (\ref{gvelsl}) is similar in form to the refractive index of a photon in Quantum Electro-Dynamics (QED) passing through a gas of charged particles, upon making the substitutions that yield the gravitational inverse square law from the corresponding Coulomb force law. More precisely, one must replace the charge density by the mass density of scatterers, set the charge per unit mass equal to $1$ and replace the constant $\sfrac{1}{4\pi\epsilon_0}$ (where $\epsilon_0$ is the permeability of the vacuum) by {\em the opposite of} the gravitational constant, namely $-G$. Note that this minus sign is crucial, in that it implies a sub-luminal group velocity for photons due to vacuum polarization effects.

Thus, for photons in a flat space-time, scattered of a density of free (non-interacting) charged particles, we may write the induced index of refraction (in natural units where $\epsilon_0=1$) as\footnote{This is obtained from the standard expression following the optical theorem, according to which the index of refraction is expressed in terms of the coherent forward scattering amplitude for a photon with polarization $\lambda$ as
\be
	n(\omega) = 1 + \frac{2\pi \,N}{k^2} f_{\lambda \lambda}(0) ~, \nonumber
\ee
where $N$ is the number density of the scatterers, $m$ denotes their mass, $k$ is the wave-vector of light (equivalently $k$ may be replaced by the frequency $\omega$ of photons assumed massless) and in the framework of a quantum field theory model
\be
	f_{\lambda \lambda}(0) = \frac{1}{8\pi m} \,{\mathcal M}_{\lambda\lambda}(k,k^\prime \rightarrow k,k^\prime) ~, \nonumber
\ee
with the overall phase of the field theory amplitude ${\mathcal M}_{\lambda\lambda}$ fixed by the optical theorem, relating the total scattering cross section to the imaginary part of $f_{\lambda\lambda}(0)$.}
\be \label{riqed}
	n^\mathrm{vac.\,pol.}_{\rm \gamma} = 1 - \frac{q^2 e^2 \,{\tilde \rho}}{2 \,m^2 \,\omega^2} + \dots
\ee
with respectively ${\tilde \rho} > 0$, $m$ and $qe$ the mass density, mass and charge of the charged carriers. The $\dots$ indicate sub-leading positive contributions coming from polarizability of the scatterer, which are either constant or proportional to positive powers of $\omega^2$.

It should be noted that the expression (\ref{riqed}) is generic and may incorporate milli-charged dark matter candidates that may exist in some models of particle physics~\cite{latimer}. If we ignore such milli-charge dark matter candidates (which do not exist in the majority of phenomenologically relevant models), then it becomes clear that the photon polarization refractive index effects are sub-leading compared to the ones induced by the scattering of photons off (neutral) dark matter, which is the dominant candidate by several orders of magnitude. For instance, the dominant source of charged scatterers in the universe are protons, for which the corresponding cosmic energy density, that is the baryon density, is two orders of magnitude smaller than the dark matter density; the $\Lambda$CDM parameters today read $\Omega_{\rm b}/\Omega_{\rm DM} \simeq 2.2 \times 10^{-2}$~\cite{Planck}.

Hence, for all practical purposes we only consider the effects on the photon refraction of the weak gravitational potential induced by the matter density $\rho_{\rm m}$, which, according to the analysis done in Ref.~\cite{Peters}, are negligible. That is, to linear order in the weak gravitational potential at hand, the refractive index of photons with low-frequencies should be considered as that of the vacuum, upon ignoring vacuum polarization effects
\be \label{photons}
	n_{\rm \gamma}^{\rm DM} \simeq 1 ~.
\ee
In this sense, at the low-frequency regime we are interested in, the photons behave as light-like particles.

%%===========================================================================%%
\subsection{Purely stringy effects of D-matter}

We should remark at this point that, in the context of the D-particles foam, there are also terms in the refractive index that scale linearly with the frequency $\omega$, which arise from the non-trivial interactions of the D-particles with the photons, viewed as open strings~\cite{delays}, that can be captured by the D-matter defects. Such terms stem from the stringy uncertainty principle, $\Delta t \,\Delta x \ge \alpha^\prime$, and can be computed by considering string scattering amplitudes of open strings, representing the photons, off a D-particle background~\cite{delays}. Taking into account the cosmic expansion, the induced delays of photons with observed frequencies $\omega$ due to these purely stringy effects are of the form~\cite{delays}
\be \label{stringydelays}
	\Delta t_{\rm o} ^{\rm D-foam} \simeq \int_0^z {\rm d}\tilde z \;C \,\frac{(N_{\rm D}^0)^{\sfrac{1}{3}}}{M_s^2} \frac{\omega}{H_0} \frac{(1 + \tilde z)}{\sqrt{\Omega_{\rm m} (1 + \tilde z)^3 + \Omega_\Lambda}} ~,
\ee
where $C < 1$ is some fudge factor, entailing information on the momentum transfer of the incident string on a D-particle during the scattering, while $N_{\rm D}^0$ is again today's D-particle three-volume density, which in principle should read $N_{\rm D} (z)$ and depend on the redshift for inhomogeneous D-particle foam models, but for our purposes here is considered $z$-independent for small $z<10$ and thus is identified with today's value. Note that this effect is also valid, in a first approximation, for close strings such as gravitons which, by hitting the D-particle, would open and attach to the brane and thus act in a similar way. This computation is thus applicable to the case of gravitons.

%%%%%%%%%%%%%%%%%%%%%%%%%%%%%%%%%%%%%%%%%%%%%%%%%%%%%%%%%%
\section{Gravitational wave phenomenology of the D-material universe \label{sec:pheno}}

In this section we shall compare the various refractive index effects (\ref{dtmass}), (\ref{dtDM}) and (\ref{stringydelays}) against the current GW phenomenology. The aim is to derive constraints on the string scale within the context of the D-material universe.

It can be readily seen that the stringy delays (\ref{stringydelays}) are sub-leading (by at least five orders of magnitude, thus negligible) compared to the $\omega^{-2}$ terms in (\ref{riqed}), for the low-frequencies we are interested in this work and the very small D-particle densities $N_{\rm D}^0$ (\ref{Nrange}) required so that the D-matter fluid acts as dark matter in the universe~\cite{emsy}. Indeed, one has
\be
	\Delta t^{\rm D-foam}_{\rm o} \lesssim 1.4 \times 10^{-1}~M_{\rm Pl}^{-1} \ll |\Delta t^{\rm DM}_{\rm o}| = 8.4 \times 10^{14}~M_{\rm Pl}^{-1} ~,
\ee
as can be seen from (\ref{dtDM}) for GW frequencies of order of 100 Hz and where we used Eq.~(\ref{NrangeNum}) to get $N_{\rm D}^0 \lesssim 6 \times 10^{-156} \,{\widetilde \xi}_0^{-2}~M_{\rm Pl}^3 \lesssim 6 \times 10^{-154}~M_{\rm Pl}^3$ with say ${\widetilde \xi}_0 \sim 0.1$, and $M_{\rm s} \gtrsim 10^{-15}~M_{\rm Pl}$.

%%===========================================================================%%
\subsection{Constraining the condensate using experimental bounds on the graviton mass}

Once the stringy effects are ignored, one is left with two competing effects on GW propagation: (i) \emph{delays} (compared to the propagation in vacuum) due to the induced graviton-mass~(\ref{dtmass}) and (ii) \emph{advances} due to the propagation of gravitons in the weak gravitational potentials induced by D-matter and dark matter distributions~(\ref{dtDM}). In principle, as already mentioned, the above effects (\ref{dtmass}) and (\ref{dtDM}) will lead to a modification of the pattern of the GW signal, due to induced dephasings of the various frequency components comprising the signal. Below we shall discuss the conditions under which the mass effects are dominant, in which the graviton group velocity would be {\em sub-luminal}.

By comparing the two cases (\ref{dtmass}) and (\ref{dtDM}), we conclude that the graviton would have a {\em sub-luminal} propagation velocity if and only if its mass is larger than a critical minimal value
\be \label{critical}
	m_{\rm G} \ge m_{\rm G}^{\rm c} = \sqrt{\frac{\rho_{\rm m}}{2 \,M_{\rm Pl}^2}} \simeq 7 \times 10^{-62}~M_{\rm Pl} \simeq 2 \times 10^{-34}~{\rm eV} ~,
\ee
where we assume the $\Lambda$CDM value given in (\ref{bench})~\cite{Planck} for the matter density.

Equations~(\ref{masscond1}), (\ref{red}), (\ref{red4b}) and (\ref{gvelsl}) lead to the following remarks:
\begin{itemize}
\item{(A)} If quantum fluctuations are sub-dominant as compared to statistical effects, mass effects dominate over the energy density induced refraction, and \emph{sub-luminal} graviton velocities in the D-material universe are attained. In such a case, the induced mass of the graviton is (in units of $M_{\rm Pl}$) of the order of the critical density of the universe, which in the current era is by several orders of magnitude smaller than the sensitivity of aLIGO/Virgo, or even pulsar timing experiments~\cite{Baskaran:2008za} which give the strongest limit to date ({\em cf.} Eq.~(\ref{massbounds}) below).

\item{(B)} If recoil quantum fluctuations are taken into account, much larger graviton masses are allowed.\footnote{Nevertheless, the stringy effects (\ref{stringydelays}), that grow linearly with the GW frequency $\omega$, are still sub-leading, for the very low-energies we consider here,  compared with the mass and refractive index effects, that are inversely proportional to the square of $\omega$.} Indeed, in such a case, the refractive effects (see Eq.~(\ref{refraction})) due to a medium of matter scatterers with density $\rho_{\rm m}$ reduce the ``{\em effective}'' mass of the graviton, to be constrained by experiments, to
\begin{align} \label{effmass}
	0 < (m_{\rm G}^{\rm eff})^2 &\equiv m_{\rm G}^2 - \frac{\rho_{\rm m}}{2 \,M_{\rm Pl}^2} \nn 
	&= \frac{1}{M_{\rm Pl}^2} \left[ \frac{M_{\rm s}^4}{6} \,\llang u_i u_j \delta^{ij} \rrang + 2 \rho_{\Lambda^{\rm vac}} - \frac{\rho_{\rm DM+b}}{2} - \frac{7}{24} \, (\tilde{\alpha} - \tilde{\alpha}_t ) \right] ~,
\end{align}
where we remind the reader that $\rho_{\rm DM+b}$ denotes any conventional matter content of the D-material universe, including both (ordinary) baryonic matter and (conventional) dark matter. Equation (\ref{effmass}) is a \emph{necessary} and \emph{sufficient} condition for positivity of $(m_{\rm G}^{\rm eff})^2$ (that is a condition for dominance of mass effects over the refractive index ones). The reader should bear in mind that in (\ref{effmass}), $2 \rho_{\Lambda^{\rm vac}} - \frac{1}{2} \,\rho_{\rm DM+b} > 0$, as a result of the $\Lambda$CDM cosmic concordance in the current era. 
\end{itemize}

Now in what follows we shall make the assumption (as a special but quite indicative case), that $0 < \tilde a < \tilde a_t$, which is required for consistency of (\ref{red4b}) if one assumes, as we do here, that
\be \label{largecond}
	M_{\rm s}^4 \,\llang u_i u_j \delta^{ij} \rrang \gg \rho_{\rm m}^{\Lambda {\rm CDM}} ~.
\ee
The positivity of the condensates is a mild assumption we make, following Ref.~\cite{odintsov}, where such quantum condensates have been argued to provide dark energy contributions. Thus, the importance of non-zero quantum condensates lies on the fact that their presence allows a much larger induced graviton-mass than the critical density of the universe. Indeed, on requiring further
\be \label{cancell}
	\tilde a_{\rm t} \sim \tilde a + 4M_{\rm s}^4 \,\llang u_i u_j \delta^{ij} \rrang ~,
\ee
we see that (\ref{red4b}) is guaranteed even with the assumption (\ref{largecond}), and hence the conclusions of Ref.~\cite{emsy} remain unchanged.

Note that the presence of the symbol $\sim$ instead of equality in (\ref{cancell}) indicates a small but non-zero difference between the {\rm l.h.s} and {\rm r.h.s.} of the above equation of order of the critical density of the universe, which is the same order as the total (observed) energy density today $\rho_{\rm total}$ (see Eq.~(\ref{rhototal})). One may solve Eq.~(\ref{cancell}) by assuming (as an indicative example) that in the current era of the universe
\be \label{example}
	\tilde a \sim M_{\rm s}^4 \llang u_i u_j \,\delta^{ij} \rrang \quad \Rightarrow \quad \tilde a_{\rm t} \sim 5 \tilde a ~,
\ee
implying that the induced effective mass of the gravitons (\ref{effmass}) can be much larger than the total energy density of the D-material fluid, namely
\be \label{effmass2}
\left[	(m_{\rm G}^{\rm eff})^2 \,M_{\rm Pl}^2 \simeq \frac{4}{3} \,\tilde a + 2 \rho_{\Lambda^{\rm vac}} - \frac{\rho_{\rm DM+b}}{2} \sim \frac{4}{3} M_s^4 \,\llang u_i u_j \delta^{ij} \rrang \quad\right] \gg \,\{ \rho_{\Lambda^{\rm vac}}, \,\rho_{\rm DM+b} \} ~,
\ee
for $\tilde a \gg \{ \rho_{\Lambda^{\rm vac}}, \rho_{\rm DM+b} \}$, assumed in Eqs.~(\ref{largecond}) and~(\ref{example}). Thus, in this example, the effective mass of the graviton is of the same order as the mass (\ref{masscond1}) induced by the dominant ``magnetic'' field condensates. It is important to remark that here one should no longer assume the range~(\ref{Nrange}), since the quantum effects are the ones responsible for ensuring the satisfaction of the upper bound~(\ref{red4b}).

The most stringent current bounds on the mass of the graviton are given by puslar timing experiments~\cite{Baskaran:2008za}, which are stronger than the bound (\ref{ligobound}) from aLIGO's direct detection of GW~\cite{LIGO}. They give
\begin{align} \label{massbounds}
	m_{\rm G}^{\rm eff} &< 8.5 \times 10^{-24}~{\rm eV} = 3.5 \times 10^{-51}~M_{\rm Pl} \quad {\rm (pulsar)} \nn
	m_{\rm G}^{\rm eff} &< 1.2 \times 10^{-22}~{\rm eV} = 5.0 \times 10^{-50}~M_{\rm Pl} \quad {\rm (aLIGO)} ~.
\end{align}
If quantum effects are ignored, in case (A) above,  the induced mass is of the order of the current critical density of the universe and hence cannot be constrained by the current limits. However, in case (B), assuming for concreteness example (\ref{example}), then (\ref{effmass2}), (\ref{massbounds}) imply
\begin{align} \label{abounds2}
	\tilde a &< 9.2 \times 10^{-102}~M^4_{\rm Pl} \quad {\rm (pulsar)} \nn
	\tilde a &< 1.9 \times 10^{-99}~M^4_{\rm Pl} \quad {\rm (aLIGO)} ~,
\end{align}
namely the upper bounds are much larger values (by several orders of magnitude) than the $\Lambda$CDM critical density.

%%===========================================================================%%
\subsection{Gravitational Cherenkov radiation}

The sub-luminal nature of the graviton in the case considered above implies other effects, independent of the GW aLIGO observations, which may constrain further the string scale in our model. We will thus investigate gravitational Cherenkov radiation~\cite{Nelson}, namely the emission of a graviton from a highly relativistic particle, propagating with velocity almost equal to that of the speed of light {\em in vacuum}. Such a process is kinematically allowed, provided the graviton group velocity is less than the speed of light {\em in vacuum}. We will therefore examine under what conditions, if at all, such an effect exists in the D-material universe. In the affirmative case, following Ref.~\cite{Nelson} and using ultra-high-energy cosmic rays data, we shall impose constraints on the lower allowed bound of the graviton propagation speed.

For electrically charged particles, the D-matter medium looks transparent~\cite{delays,sakharov}, on account of gauge invariance properties. This is the case of the ultra-high-energy cosmic rays (UHECR), which therefore can propagate in the D-matter medium, for which (\ref{critical}) is satisfied, with a speed higher than that of (low-frequency) gravitons, and therefore gravitational Cherenkov radiation is kinematically allowed~\cite{Nelson}. As a result, cosmic rays will loose energy. The observation of the most energetic cosmic rays, with energies $10^{20}~{\rm eV}$, implies then stringent constraints on the lower bound of the propagation velocity of such sub-luminal low-frequency gravitons. According to the analysis in Ref.~\cite{Nelson} and depending on the assumptions on the origin (galactic or extragalactic) of the UHECR, one obtainss the bounds
\begin{align} \label{cherenkov}
	0 &< 1 - v_{\rm g} < 2 \times 10^{-15} \qquad \mathrm{for~UHECR~of~galactic~origin} \nn
	0 &< 1 - v_{\rm g} < 2 \times 10^{-19} \qquad \mathrm{for~UHECR~of~extragalactic~origin}
\end{align}
in units of the speed of light in vacuum. From (\ref{gvelg}), upon substituting $m_{\rm G}$ by $m_{\rm G}^{\rm eff}$ (\ref{effmass}), we then obtain the bounds
\begin{align} \label{uhecr}
	(m_{\rm G}^{\rm eff})^2 &< 4 \times 10^{-15} \;\omega^2 ~ \quad \mathrm{for~UHECR~of~galactic~origin} \nn
	(m_{\rm G}^{\rm eff})^2 &< 4 \times 10^{-19} \;\omega^2 ~ \quad \mathrm{for~UHECR~of~extragalactic~origin}
\end{align}
which, in the example (\ref{example}) leading to (\ref{effmass2}) and for the frequency range $\omega \simeq 100~{\rm Hz} \simeq 1.7 \times 10^{-40}~M_{\rm Pl}$ of the GW of aLIGO~\cite{LIGO}, yields
\begin{align} \label{uhecr2}
	\tilde a &< 8.7 \times 10^{-95}~M_{\rm Pl}^4 ~ \quad \mathrm{for~UHECR~of~galactic~origin} \nn
	\tilde a &< 8.7 \times 10^{-99}~M_{\rm Pl}^4 ~ \quad \mathrm{for~UHECR~of~extragalactic~origin}~.
\end{align}
Thus, if UHECR are of extragalactic origin, then the bounds on the minimal value of the (sub-luminal) graviton propagation speed obtained as a consequence of the gravitational Cherenkov radiation, are at best of the same order of magnitude as the bounds 
(\ref{massbounds}), otherwise (namely, for UHECR of galactic origin) the corresponding bounds are several orders of magnitude weaker.

\section{Conclusions} 

In this work, which builds upon our previous studies of the phenomenology of the D-material universe~\cite{emsy,msy}, we considered the effects of the recoiling D-particles on the propagation of GW. We have argued that, for the low-energy regime of interest for GWs observed by aLIGO, which was the focus of our attention here,  the main effect is an induced effective mass for the graviton, given by (\ref{effmass2}), which, depending on the magnitude of the 
D-particle recoil-velocity fluctuations, can be much larger than the vacuum energy and dark matter density, and hence can be bounded by 
pulsar timing or aLIGO  measurements.  As the magnitude of such quantum fluctuations cannot be determined theoretically at present, due to uncertainties in the underlying dynamics of the collection of D-particle defects that require going beyond the current perturbative analysis in brane/string theory, such studies can only be phenomenological at present, and this is what we concentrated upon in this work.

One of the most important features of a massive graviton is that it is \emph{sub-luminal} as compared to photons, leading to a negligible refractive index effect for the low-energy regime of interest to us here. In that case,  gravitational Cherenkov radiation may impose additional constraints, in particular if one considers ultra-high-energy cosmic rays of an extragalactic origin. As we have shown,  one gets upper bounds on graviton masses comparable (in order of magnitude) to those obtained from aLIGO interferometric measurements of GW, but still weaker than those obtained from pulsar timing data.
Certainly Cherenkov radiation bounds may improve in the future  (once higher energies can be probed), although it must be said that, if the GZK cutoff of the highest allowed energy of UHECR remains valid, then the current bounds  (\ref{uhecr2}) may not change significantly. (We remind the reader that a D-particle medium is assumed here virtually transparent to the UHECR, in view of their electrical charge.) Of course if Lorentz violation due to the D-particle populations is significant, then the optical transparency of the high-energy universe may be further affected, modifying the above discussed bounds. 

Another important comment we feel stressing once more, concerns the point of view taken in this work as far as the effects of recoil-velocity condensates on the GW propagation are concerned. Any such effects contribute to the so-called vacuum energy (of a type that is approximately a cosmological constant during the current epoch of the universe). In the context of our D-material universe,  such contributions were assumed as being due to the stress tensor of string matter rather than the geometry, that is they are due to the right-hand side of the corresponding (low-energy) Einstein equations, rather than the left-hand one. This allowed us to treat any such effect as corrections on top of a virtually flat-space-time Minkowski background, where a mass for the graviton can be defined. If however, one used de Sitter (or anti-de Sitter) space-time as the maximally symmetric backgrounds, around which metric fluctuations are considered, then cosmological constant (including recoil-velocity condensate) effects are not interpreted as graviton mass, except perhaps in the anti-de Sitter case, where such effects amount to a cosmological constant $\Lambda < 0$ and the graviton mass is of the order $|\Lambda|^{1/2}$. 
Unfortunately, such important issues can only be resolved when a full dynamical microscopic theory of the underlying D-particle dynamics in the D-material universe is available. At present, we are far from such a detailed description. Nevertheless, we hope that the D-material universe as a concept is an interesting one, especially because it seems that the model is capable of passing all of the current phenomenological challenges, including the effects of the medium on the GW.

%%%%%%%%%%%%%%%%%%%%%%%%%%%%%%%%%%%%%%%%%%%%%%%%%%%%%%%%%%%%%%%%%%
\vspace{1em} 
\section*{Acknowledgments}

The work of T.E. is supported by a KCL GTA studentship, and that of N.E.M. is partially supported by the London Centre for Terauniverse Studies, using funding from the ERC Advanced Investigator Grant 267352, and by STFC (UK) under the research grant ST/L000326/1. We acknowledge discussions with members of the LIGO Scientific Collaboration and the Work Package 3 of the Theory Working Group of the Euclid Consortium, to both of which M.S. is a member.

%%%%%%%%%%%%%%%%%%%%%%%%%%%%%%%%%%%%%%%%%%%%%%%%%%%%%%%%%%%%%%%%%%

\end{document}